\documentclass[twoside,twocolumn,english,prl,a4paper,elide,showpacs,superscriptaddress]{revtex4}
\usepackage[T1]{fontenc}
\usepackage[latin9]{inputenc}
\usepackage[a4paper]{geometry}
\usepackage{natbib}
\usepackage{textcomp}
\usepackage{relsize}
\usepackage{graphicx}
\usepackage{amssymb}
\usepackage{amsmath}
\usepackage{color}

\makeatletter

\@ifundefined{textcolor}{}
{%
 \definecolor{BLACK}{gray}{0}
 \definecolor{WHITE}{gray}{1}
 \definecolor{RED}{rgb}{1,0,0}
 \definecolor{GREEN}{rgb}{0,1,0}
 \definecolor{BLUE}{rgb}{0,0,1}
 \definecolor{CYAN}{cmyk}{1,0,0,0}
 \definecolor{MAGENTA}{cmyk}{0,1,0,0}
 \definecolor{YELLOW}{cmyk}{0,0,1,0}
 }

\makeatother

\usepackage{babel}

\begin{document}

\preprint{This line only printed with preprint option}

\title{Measurement of Spin Projection Noise in Broadband
Atomic Magnetometry}

\author{M. Koschorreck}
    \email{marco.koschorreck@icfo.es}
    \homepage{http://icfo.es}
    \affiliation{ICFO-Institut de Ciencies Fotoniques, 08860 Castelldefels (Barcelona), Spain}

\author{M. Napolitano}
    \affiliation{ICFO-Institut de Ciencies Fotoniques, 08860 Castelldefels (Barcelona), Spain}

\author{B. Dubost}
    \affiliation{ICFO-Institut de Ciencies Fotoniques, 08860 Castelldefels (Barcelona), Spain}
   \affiliation{Laboratoire Mat\'{e}riaux et Ph\'{e}nom\`{e}nes Quantiques, Universit\'{e} Paris Diderot et CNRS, \\UMR 7162, B\^{a}t. Condorcet, 75205 Paris Cedex 13, France}
\author{M. W. Mitchell}
  \affiliation{ICFO-Institut de Ciencies Fotoniques, 08860 Castelldefels (Barcelona), Spain}

\newcommand{\mtext}[1]{{\color{blue}#1}}
\newcommand{\rtext}[1]{{\color{red}#1}}
\definecolor{orange}{rgb}{.75,0.55,.025}
\newcommand{\movtext}[1]{{\color{orange}#1}}
\newcommand{\modtext}[1]{{\color{black}#1}}
\definecolor{darkgreen}{rgb}{0,0.5,0}
\newcommand{\oldtext}[1]{{\color{darkgreen}#1}}
\newcommand{\otext}[1]{\oldtext{#1}}
\newcommand{\qtext}[1]{{\color{cyan}#1}}

\newcommand{\expect}[1]{\left<#1\right>}
\newcommand{\var}{{\mathrm{var}}}
\newcommand{\dexpect}[1]{\left\langle\right.#1\left.\right\rangle}

\begin{abstract}
We measure the sensitivity of a broadband atomic magnetometer
using quantum non-demolition spin measurements. A cold,
dipole-trapped sample of rubidium atoms provides a long-lived spin
system in a non-magnetic environment, and is probed non-destructively by paramagnetic Faraday rotation. The
calibration procedure employs a known reference state, the
maximum-entropy or `thermal' spin state and quantitative
imaging-based atom counting to identify electronic, quantum, and
technical noise in both the probe and spin system. The measurement
achieves sensitivity $2.8\,$dB better than the projection noise level ($6\,$dB better if optical noise is suppressed) and will enable squeezing-enhanced broadband magnetometry [Geremia, \emph{et al.} PRL $\mathbf{94}$, 203002 (2005)].
\end{abstract}
\maketitle

Precision magnetic field measurements can be made by optically
detecting the Larmor precession produced in a spin-polarized
atomic sample \cite{Budker2007NPv3p227}. The technique is ultimately
limited by quantum noise, present in both the optical measurement
and in the atomic system itself.  Recent works using
large numbers of atoms and long spin coherence times have
demonstrated sub-fT/$\sqrt{{\mathrm {Hz}}}$ sensitivities for DC
\cite{Kominis2003Nv422p596} and RF \cite{Wasilewski2009Avquant-php}
fields for bandwidths of order $1\,$kHz, surpassing superconducting
sensors (SQUIDS) in sensitivity and approaching quantum noise limits.
Potential applications of magnetic sensors range
from gravitational-wave detection \cite{Harry2000APLv76p1446}
 to magnetoencephalography \cite{Hamalainen1993RMPv65p413}.

Atomic spin readout using optical quantum non-demolition (QND) measurement \cite{Braginsky1974UFNv114p41,Grangier1992OCv89p99} allows magnetometry to
surpass the standard quantum limit $\delta B \propto 1/\sqrt{N}$ associated with atomic projection noise \cite{Kuzmich1999PRAv60p2346}.  Similarly, optical squeezing can surpass the shot-noise limit in optical measurements \cite{Hetet2007JoPBv40p221, Predojevic2008PRAv78p63820}.  The measurement is then constrained by the much weaker Heisenberg limit $\delta B \propto 1/N$.  This strategy is particularly well adapted to broadband magnetometry, in which repeated or continuous measurements determine a time-varying field.  Each QND measurement both indicates the measured spin variable and (ideally) projects the system onto a spin-squeezed state, increasing the sensitivity of subsequent measurements.  To date, QND probing of spin variables has achieved projection-noise limited precision only on magnetically insensitive "clock" transitions \cite{Appel2009PNASv106p10960, Schleier-Smith2009Avp}.  A significant obstacle has been, up to now, the calibration of the spin noise measurements in a magnetically sensitive system \citep{Geremia2004Sv304p270,*Geremia2005PRLv94p203002,*Geremia2008PRLvp1}.

  We report here a cold, trapped atomic ensemble with a spin lifetime
of up to 30 seconds, a spin measurement bandwidth of 1 MHz, and a
spin readout noise of approximately 500 spins, 2.8 dB below the projection noise level.
Optical shot noise accounts for most of the remaining noise. Recent experiments on atom-tuned squeezed light show a reduction of light noise by $5\,$dB \cite{Hetet2007JoPBv40p221}, which would reduce readout noise further, to  $6\,$dB below projection noise.
We establish the projection noise level by two techniques: a
calibrated measurement of the per-atom optical rotation, and an
analysis of noise scaling when measuring a reference state.  The
use of noise reference states, e.g. thermal, vacuum, or coherent
states, is well-established in quantum optics.  To extend this to
spin systems, we use the maximum-entropy state, also known as the
`thermal' spin state.

The experiments are performed with a macroscopic sample of $^{87}$Rb atoms held in an optical dipole trap. After laser cooling, atoms are loaded into the weakly-focused beam of a Yb:YAG laser at $1030\,\mathrm{nm}$. The sample contains about one million atoms at temperatures of about $25\mu$K. Tight (weak) confinement in the transverse (longitudinal) direction produces a sample with high aspect ratio of  ${\simeq240:1}$.  This geometry produces a large atom-light interaction for light propagating along the trap axis.  In earlier experiments, we have measured an effective on-resonance optical depth of above 50 \citep{Kubasik2009PRAv79p43815}.

\begin{figure*}[!]
\includegraphics[width=0.9\textwidth]{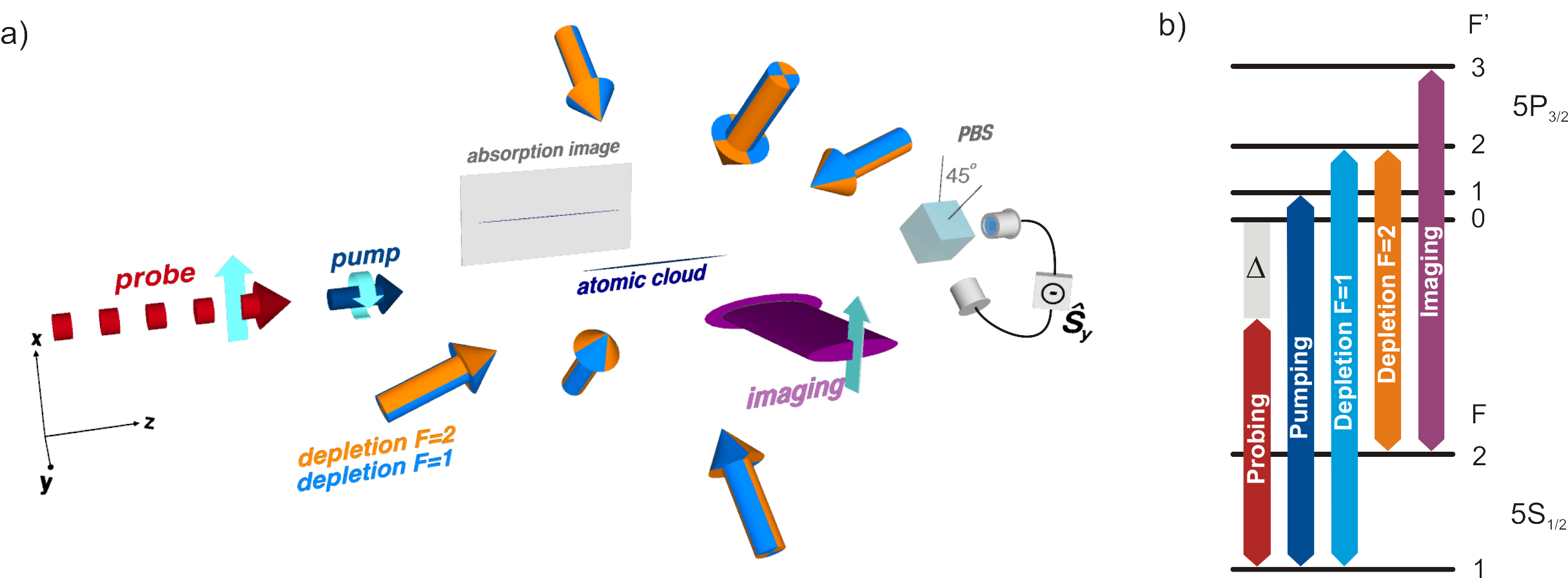}

\caption{\label{fig:setup}(Color online) a) Atomic ensemble with probing, pumping, and imaging light fields. The polarimeter measures in the $45^\circ$-basis, i.e., the Stokes component $\hat{S}_y$; b) Atomic transitions for probing, preparation, and imaging light fields.}

\end{figure*}

\newcommand{\measname}{{fast $N_A$}}

The collective spin is measured using paramagnetic Faraday
rotation with an off-resonance probe.  The ensemble spin,
$\hat{\bf F}$, interacts with an optical pulse of duration $\tau$
and polarization described by the vector Stokes operator $\hat{\bf S}$ through the effective Hamiltonian \cite{Madsen2004PRAv70p52324}
\begin{equation}
   \hat{H}=\hbar\frac{G}{\tau}\hat{S}_{z}\hat{F}_{z}\,\,.\label{eq:H_int}
\end{equation}
We define $\hat{\bf S}$ in terms of annihilation (creation) operators for left and right circularly polarized light modes, $\hat{a}_{\pm}$($\hat{a}^{\dag}_{\pm}$) , as $\hat{S_i}\equiv(\hat{a}^{\dag}_+\hat{a}^{\dag}_-)\sigma_i(\hat{a}_+\hat{a}_-) $ \cite{Jauch1976}, where $\sigma_i$ are Pauli matrices.
The interaction strength $G$ depends on transition dipole moments,
optical detuning, and beam and atom cloud geometry
\citep{Geremia2006PRAv73p42112}.

A light pulse experiences
the polarization rotation (to first order in $H$)
\begin{equation}
\hat{S}_{y}^{(\mathrm{out})}=\hat{S}_{y}^{(\mathrm{in})}+ G\hat{
S}_{x}^{(\mathrm{in})}\hat{F}_{z}^{(\mathrm{in})}\label{eq:SyOut}
\end{equation} where superscripts $(\mathrm{in}),(\mathrm{out})$ indicate components before and after the interaction, respectively.  In a
QND measurement of $\hat{F}_z$, the input state has $\dexpect{\hat{S}_{x}}=N_{L}/2$ and $\dexpect{\hat{S}_{y}}=\dexpect{\hat{S}_{z}}=0$ such that $\hat{F}_z$ can be estimated as
$\hat{F}_{z}^{(\mathrm{in})} \approx
2\hat{S}_{y}^{(\mathrm{out})}/G N_L$.  In addition, {macroscopic} rotations can be used to measure $N_A$, by polarizing the ensemble such that
$\dexpect{\hat{F}_z} = N_A\,$  prior to probing.  We refer to this as a {``dispersive'' atom number} measurement and calibrate it using quantitative absorption
imaging.

To establish the sensitivity at the quantum level, we note that for input states with ${\dexpect{\hat{S}_{y}^{(\mathrm{in})}}=\dexpect{\hat{F}_{z}^{(\mathrm{in})}}=0}$,
without initial correlation between  $\hat{S}_{y}^{(\mathrm{in})}$
and $\hat{F}_{z}^{(\mathrm{in})}$, and
$\mathrm{var}(\hat{S}_{x})\ll \dexpect{\hat{S}_{x}}^2$, the
polarization variance is
\begin{equation}
    \mathrm{var}(\hat{S}_{y}^{(\mathrm{out})})=\mathrm{var}
(\hat{S}_{y}^{(\mathrm{in})})+
G^{2}\frac{N_{L}^{2}}{4}\mathrm{var}(\hat{F}_{z}^{(\mathrm{in})})\,\,\label{eq:varSy}.
\end{equation}

The first term, the input optical polarization, in general has
variance $\var(\hat{S}_{y}^{(\mathrm{in})}) = N_L/4+\alpha
N_L^2$, where the part proportional to $N_L$ indicates intrinsic noise and
 the one proportional to $N_L^2$ indicates technical noise due to variations in the
optical state preparation. Similarly, the second term contributes
a variance $G^2 (\hat{S}_{x}^{(\mathrm{in})})^2 \var(\hat{F}_z)$ with
$\var(\hat{F}_z) = N_A V_1+\beta N_A^2 V_1$ where $V_1$ is the
variance per atom.  
Finally, we must add a constant ``electronic noise'' $V_E$ from the detector, and arrive to the measurable signal

\begin{eqnarray}
\mathrm{var}(\hat{S}_{y}^{(\mathrm{meas})})&=& V_E +
\frac{N_L}{4} + \alpha N_L^2 + G^2 V_1 \frac{N_L^2}{4} N_A \nonumber \\ &&+ \beta G^2 V_1 \frac{N_L^2}{4} N_A^2.
\label{Eq:Scalings}
\end{eqnarray}

 Equation (\ref{Eq:Scalings}) contains the essential elements of
the calibration technique.  All terms have distinct scaling with photon and atom number, and can thus be separately identified if
$\mathrm{var}(\hat{S}_{y}^{(\mathrm{meas})})$ is measured as a
function of $N_L$ and $N_A$. The terms in $N_L$ and $N_L^2 N_A$
correspond to quantum noise of light and atoms, respectively.
Together they provide an absolute calibration of the gain of the
detection system and the atom-light coupling $G$. The remaining
terms represent various noise sources.  Only if these are
simultaneously small relative to the atomic quantum noise, 
quantum signals will be detectable.

For $N_A$ atoms with spin quantum number $F$, the reference state
is $\rho = \rho_T^{\otimes N_A}$, where $\rho_T$ is the
completely-mixed state of dimension $2F+1$. In terms of the
collective spin $\hat{\bf F} \equiv \sum_i \hat{\bf f}^{(i)}$
where $\hat{\bf f}^{(i)}$ is the spin of the $i$-th atom, the
thermal state has zero average value, and a noise of
$\var(\hat{F}_n) = \frac{1}{3} F(F+1) N_A$, where $\hat{F}_n$ is
any spin component. Hence $V_1 = \frac{1}{3} F(F+1)$.

We now describe in detail the experimental methods. For each pulse, the
photon number $N_L$ is measured by: splitting off a portion of the probe  beam before it propagates through the atoms, detection with a calibrated photodiode, and numerical integration of the waveform.
Absolute measurement of $N_A$ is carried out by quantitative
absorption imaging \citep{Lewandowski2003JoLTPv132p309,
Ketterle1999PISoPFvp67}: atoms are transferred into the $F=2$ hyperfine
ground state by $100\,\mu\mathrm{s}$ of laser light tuned to the
${F=1\rightarrow F'=2}$ transition.  The dipole trap is switched off
to avoid spatially-dependent light shifts and an image is taken
with a $100\,\mu\mathrm{s}$ linearly-polarized pulse resonant to
the ${F=2\rightarrow F'=3}$ transition. A background image is taken
under the same conditions, but without atoms. The observed $N_A$ error is $<4\%$ (RMS) including loading fluctuations and measurement noise. The measurement noise is thus well below $4\%$.

For fast and non-destructive $N_A$ determination, we use {dispersive atom-number}  measurement: the sample is spin-polarized along $z$ by on-axis
optical pumping with $50\,\mu\mathrm{s}$ of circularly-polarized
light tuned to the $F=1\rightarrow F'=1$ transition. At the same
time, light resonant to the ${F=2\rightarrow F'=2}$ transition (via
the MOT beams) prevents accumulation of atoms in $F=2$. We define a quantization axis by applying a small bias field of $\sim100\,$mG along $z$. Probe
pulses, tuned $800\,\mathrm{MHz}$ to the red of the
${F=1\rightarrow F'=0}$ transition are used to measure the rotation
angle $\phi = N_A G$ with $N_A$ measured by absorption imaging immediately afterward. We find $G=6.6(5)\times10^{-8}$.

Thermal spin states for atoms in the $F=1$ manifold are produced
by repeatedly optically pumping atoms from ${F=1\rightarrow F=2}$
and back, using lasers tuned to the ${F=1\rightarrow F'=2}$ and
${F=2\rightarrow F'=2}$ transitions, and applied from six different
directions.  Each pumping cycle takes $300\,\mu$s.  To avoid any
residual polarization, we apply bias fields of ${B_{z}=135\,}$mG,
${B_{y}=140\,}$mG, and ${B_{x}=270\,}$mG, respectively during the three
back-and-forth cycles.  Finally, the $F=2$ manifold is further
depleted with $100\,\mu s$ of resonant light on the
${F=2\rightarrow F'=2}$ transition with
zero magnetic field.  After these steps, no remaining mean
polarization along $z$ is observed. This procedure is designed to transfer disorder from the thermalized center-of-mass degrees of freedom to the spin state: Illumination from
six directions produces a polarization field with sub-wavelength
structure, in which the atoms are randomly distributed. Possible net imbalances in the pump polarizations are scrambled by the application of different bias fields.

The measurement of $\hat{F}_z$ is made by sending a train of $1\,\mu$s long pulses with  $10\,\mu$s period to the atoms.
Each pulse contains about $25\times10^6$ photons,
vertically polarized and tuned $800\,\mathrm{MHz}$ to the red of
the ${F=1\rightarrow F'=0}$ transition.  The output pulses are
analyzed in the $\pm 45^{\circ}$ basis with an ultra-low-noise
balanced photo-detector \citep{Windpassinger2009MSTv20p55301}, giving a direct measure of  $\hat{S}_y$.  This signal,
as well as the signal of the photon-number reference detector, are
recorded on a digital storage oscilloscope for later evaluation.
While it is possible to vary $N_L$ by adjusting the probe
power or pulse duration, it is more convenient to
sum the signals from multiple pulses in ``meta-pulses,'' containing a larger total number of photons. As we are in the linear regime, a meta-pulse will have the same information as a single higher-energy pulse.

The projection noise measurement proceeds as follows: the
dipole trap is loaded (3s) and we wait $400\,\mathrm{ms}$ to allow
motional thermalization and the escape of untrapped atoms. We then repeat the following sequence 20 times: preparation of a thermal spin state,
QND measurement of $\hat{F}_z$, and dispersive $N_A$ measurement.  In
each cycle $\approx 15\%$ of the atoms are lost from the dipole
trap, mostly during state preparation, so that different values of
$N_A$ are sampled during the measurement sequence. The entire sequence is repeated 500 times to acquire statistics. In a separate experiment, under the same conditions, we measure the coupling constant $G$, using parametric Faraday rotation by a $z$-polarized
sample and absorption imaging as described above.

\begin{figure}
\begin{centering}
\includegraphics[width=1\columnwidth]{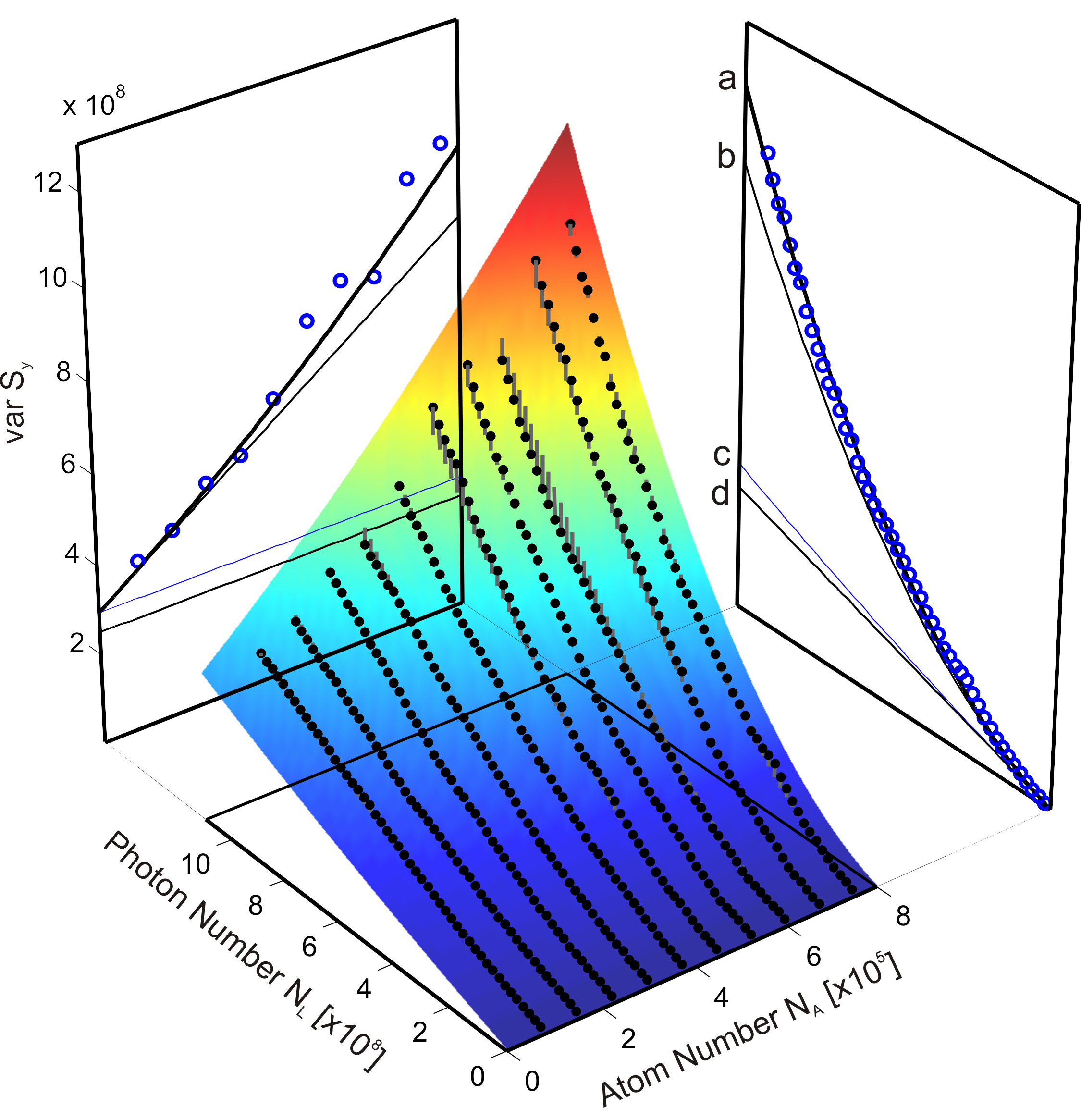}
\par\end{centering}

\caption{\label{fig:PN}(color online) Measured variance of $\hat{S}_y$ as black dots
and a fit to the data using Eq. (\ref{Eq:Scalings}) as colored surface.  The left plot shows
 $\var(\hat{S}_y)$ vs. atom number for  ${N_L = 10^9}$ photons. See Fig. \ref{fig:PNvsNA} for more details. The right plot shows 
 $\var(\hat{S}_y)$ vs. photon
number for ${N_A = 7.6 \times 10^5}$ atom numbers. In the left and right plot curves indicate: a) total noise, b) projection noise plus light noise, c) light shot and technical noise, and d) light shot noise.}

\end{figure}

 Experimental data for atom numbers
between $4 \times 10^4$ and $8 \times 10^5$ and photon numbers up
to $10^9$ are shown in Figure \ref{fig:PN}. The data are fitted with the theoretical
expression (\ref{Eq:Scalings}) which is shown as a surface. The
deduced coupling constant is ${G=6.65(3)\times10^{-8}}$ and the electronic noise level ${V_E=4.9\times10^5}$. The coefficients for the technical noise are ${\alpha=4.3(1)\times10^{-11}}$ and ${\beta=3.1(7)\times10^{-7}}$.
Atomic projection noise dominates for a large range of $N_L$ and $N_A$ above other technical and quantum noise sources, as seen in the vertical panels of Fig. \ref{fig:PN}.

\begin{figure}
\begin{centering}
\includegraphics[width=1\columnwidth]{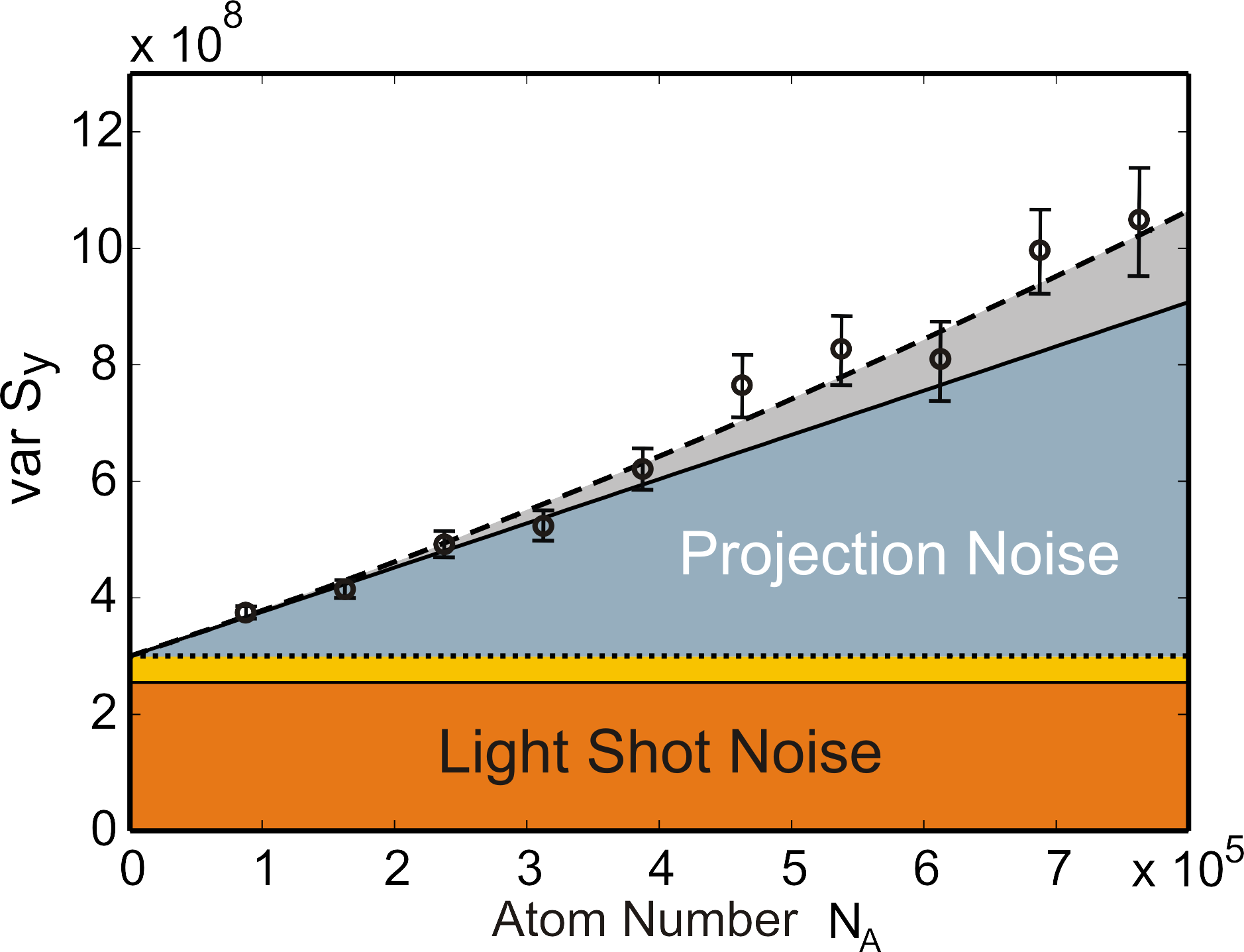}
\par\end{centering}

\caption{\label{fig:PNvsNA}(Color online) Measured variance of $\hat{S}_y$ with statistical errors for ${N_L=10^{9}}$ as a function of atom-number. Dashed curve: Theoretical curve including technical noise sources. Solid Line: Pure
spin projection noise. Dotted Line: Shot noise and technical light noise, Thin solid line: Light shot noise. The electronic noise is not plotted because it is negligible for this number of photons.}
\end{figure}

For the maximum number of photons ${N_L=1\times10^9}$, the noise scaling with atom number is highlighted in Fig. \ref{fig:PNvsNA}.  For
the largest atom number measured, i.e., ${N_A = 7.6 \times 10^5}$,
the light shot noise, atomic technical noise, light
technical noise and electronic noise are  $3.5\,$dB, $6.3\,$dB, $11.2\,$dB and $30\,$dB  below the projection noise level, respectively. At this point, the projection noise corresponds to $\var{(\hat{F}_z})^{1/2}\sim700$ spins.

The two independent measurements of $G$, by noise scaling with a thermal state and by macroscopic rotation with a polarized state, agree to within statistical uncertainties of less than ten percent.  Systematic errors due to imperfect state preparation are considerably below this: Errors in preparation of the thermal state are observed to be below the projection noise level, i.e., less than parts-per-thousand RMS, both in average value $\dexpect{\hat{F}_z}$ and in the technical noise shown in Figure \ref{fig:PNvsNA}. 
 For the polarized state measurement of $G$, simulation indicates that $>99.9\%$ polarization can be achieved with our pumping power and duration.  Furthermore, if this pumping is imperfect, it leads to an underestimate of the signal-to-noise ratio: smaller $\dexpect{\hat{F}_z}$ would cause less rotation and an underestimate of $G$.

The light technical noise may be due to small
imbalance of the polarization analyzer and thermal birefringence
produced by the dipole laser. Active stabilization of the balancing could improve and reduce the light technical noise considerably.
Atomic technical noise may come from classical fluctuations in the lasers during optical pumping.

Extrapolating the technical noise of atoms and light, both remain below their respective quantum noise terms up to $N_{A,\mathrm{qn}}\equiv\beta^{-1}=3.2\times10^6$ and
$N_{L,\mathrm{qn}}\equiv (4\alpha)^{-1}=5.8\times10^9$, respectively.  It would thus be possible to increase the number of atoms in the trap while remaining projection-noise limited.

In summary, we have demonstrated sub-projection noise sensitivity of QND spin measurement  in a broad-band atomic magnetometer.  Unlike previous attempts, we use noise scaling and a thermal state to obtain an absolute quantification of the measurement noise.  The results are confirmed by independent quantification of the QND measurement gain, i.e., the atom-light interaction strength. 
The new method detects different noise sources, i.e., atomic and light quantum and technical noise, and the electronic noise floor, by their respective scaling with atom and photon number.  
The results indicate that it will be possible to increase the sensitivity of magnetometers with MHz-bandwidth by applying measurement induced squeezing. This can have important implications for spatially resolved magnetometry, where cold atomic systems have demonstrated $\mu$m-resolution \cite{Aigner2008Sv319p1226}. Also in the field of quantum information processing, projection-noise limited QND measurements play an essential role for quantum memory and quantum cloning tasks \cite{Echaniz2008PRAv77p32316}.

We thank Robert Sewell for careful reading of the manuscript.  This work was funded by the Spanish Ministry of Science and Innovation under the ILUMA project (Ref. FIS2008-01051) and the Consolider-Ingenio 2010 Project \textquotedblleft{}QOIT\textquotedblright{}.

\bibliographystyle{apsrev}
\bibliography{PNthermArX.bbl}

\begin{thebibliography}{24}
\expandafter\ifx\csname natexlab\endcsname\relax\def\natexlab#1{#1}\fi
\expandafter\ifx\csname bibnamefont\endcsname\relax
  \def\bibnamefont#1{#1}\fi
\expandafter\ifx\csname bibfnamefont\endcsname\relax
  \def\bibfnamefont#1{#1}\fi
\expandafter\ifx\csname citenamefont\endcsname\relax
  \def\citenamefont#1{#1}\fi
\expandafter\ifx\csname url\endcsname\relax
  \def\url#1{\texttt{#1}}\fi
\expandafter\ifx\csname urlprefix\endcsname\relax\def\urlprefix{URL }\fi
\providecommand{\bibinfo}[2]{#2}
\providecommand{\eprint}[2][]{\url{#2}}

\bibitem[{\citenamefont{Budker and Romalis}(2007)}]{Budker2007NPv3p227}
\bibinfo{author}{\bibfnamefont{D.}~\bibnamefont{Budker}} \bibnamefont{and}
  \bibinfo{author}{\bibfnamefont{M.}~\bibnamefont{Romalis}},
  \bibinfo{journal}{Nature Physics} \textbf{\bibinfo{volume}{3}},
  \bibinfo{pages}{227} (\bibinfo{year}{2007}).

\bibitem[{\citenamefont{Kominis et~al.}(2003)\citenamefont{Kominis, Kornack,
  Allred, and Romalis}}]{Kominis2003Nv422p596}
\bibinfo{author}{\bibfnamefont{I.}~\bibnamefont{Kominis}},
  \bibinfo{author}{\bibfnamefont{T.}~\bibnamefont{Kornack}},
  \bibinfo{author}{\bibfnamefont{J.}~\bibnamefont{Allred}}, \bibnamefont{and}
  \bibinfo{author}{\bibfnamefont{M.}~\bibnamefont{Romalis}},
  \bibinfo{journal}{Nature} \textbf{\bibinfo{volume}{422}},
  \bibinfo{pages}{596} (\bibinfo{year}{2003}).

\bibitem[{\citenamefont{Wasilewski et~al.}(2009)\citenamefont{Wasilewski,
  Jensen, Krauter, Renema, and Polzik}}]{Wasilewski2009Avquant-php}
\bibinfo{author}{\bibfnamefont{W.}~\bibnamefont{Wasilewski}},
  \bibinfo{author}{\bibfnamefont{K.}~\bibnamefont{Jensen}},
  \bibinfo{author}{\bibfnamefont{H.}~\bibnamefont{Krauter}},
  \bibinfo{author}{\bibfnamefont{J.~J.} \bibnamefont{Renema}},
  \bibnamefont{and} \bibinfo{author}{\bibfnamefont{E.~S.}
  \bibnamefont{Polzik}}, \bibinfo{journal}{ArXiv}
  \textbf{\bibinfo{volume}{quant-ph}} (\bibinfo{year}{2009}),
  \eprint{0907.2453v2}.

\bibitem[{\citenamefont{Harry et~al.}(2000)\citenamefont{Harry, Jin, Paik,
  Stevenson, and Wellstood}}]{Harry2000APLv76p1446}
\bibinfo{author}{\bibfnamefont{G.~M.} \bibnamefont{Harry}},
  \bibinfo{author}{\bibfnamefont{I.}~\bibnamefont{Jin}},
  \bibinfo{author}{\bibfnamefont{H.~J.} \bibnamefont{Paik}},
  \bibinfo{author}{\bibfnamefont{T.~R.} \bibnamefont{Stevenson}},
  \bibnamefont{and} \bibinfo{author}{\bibfnamefont{F.~C.}
  \bibnamefont{Wellstood}}, \bibinfo{journal}{Appl. Phys. Lett.}
  \textbf{\bibinfo{volume}{76}}, \bibinfo{pages}{1446} (\bibinfo{year}{2000}).

\bibitem[{\citenamefont{H\"am\"al\"ainen
  et~al.}(1993)\citenamefont{H\"am\"al\"ainen, Hari, Ilmoniemi, Knuutila, and
  Lounasmaa}}]{Hamalainen1993RMPv65p413}
\bibinfo{author}{\bibfnamefont{M.}~\bibnamefont{H\"am\"al\"ainen}},
  \bibinfo{author}{\bibfnamefont{R.}~\bibnamefont{Hari}},
  \bibinfo{author}{\bibfnamefont{R.~J.} \bibnamefont{Ilmoniemi}},
  \bibinfo{author}{\bibfnamefont{J.}~\bibnamefont{Knuutila}}, \bibnamefont{and}
  \bibinfo{author}{\bibfnamefont{O.~V.} \bibnamefont{Lounasmaa}},
  \bibinfo{journal}{Rev. Mod. Phys.} \textbf{\bibinfo{volume}{65}},
  \bibinfo{pages}{413} (\bibinfo{year}{1993}).

\bibitem[{\citenamefont{Braginsky and
  Vorontsov}(1974)}]{Braginsky1974UFNv114p41}
\bibinfo{author}{\bibfnamefont{V.}~\bibnamefont{Braginsky}} \bibnamefont{and}
  \bibinfo{author}{\bibfnamefont{Y.}~\bibnamefont{Vorontsov}},
  \bibinfo{journal}{Usp Fiz. Nauk} \textbf{\bibinfo{volume}{114}},
  \bibinfo{pages}{41} (\bibinfo{year}{1974}).

\bibitem[{\citenamefont{Grangier et~al.}(1992)\citenamefont{Grangier, Courty,
  and Reynaud}}]{Grangier1992OCv89p99}
\bibinfo{author}{\bibfnamefont{P.}~\bibnamefont{Grangier}},
  \bibinfo{author}{\bibfnamefont{J.}~\bibnamefont{Courty}}, \bibnamefont{and}
  \bibinfo{author}{\bibfnamefont{S.}~\bibnamefont{Reynaud}},
  \bibinfo{journal}{Opt. Commun.} \textbf{\bibinfo{volume}{89}},
  \bibinfo{pages}{99} (\bibinfo{year}{1992}).

\bibitem[{\citenamefont{Kuzmich et~al.}(1999)\citenamefont{Kuzmich, Mandel,
  Janis, Young, Ejnisman, and Bigelow}}]{Kuzmich1999PRAv60p2346}
\bibinfo{author}{\bibfnamefont{A.}~\bibnamefont{Kuzmich}},
  \bibinfo{author}{\bibfnamefont{L.}~\bibnamefont{Mandel}},
  \bibinfo{author}{\bibfnamefont{J.}~\bibnamefont{Janis}},
  \bibinfo{author}{\bibfnamefont{Y.~E.} \bibnamefont{Young}},
  \bibinfo{author}{\bibfnamefont{R.}~\bibnamefont{Ejnisman}}, \bibnamefont{and}
  \bibinfo{author}{\bibfnamefont{N.~P.} \bibnamefont{Bigelow}},
  \bibinfo{journal}{Phys. Rev. A} \textbf{\bibinfo{volume}{60}},
  \bibinfo{pages}{2346} (\bibinfo{year}{1999}).

\bibitem[{\citenamefont{Hétet et~al.}(2007)\citenamefont{Hétet, Glöckl,
  Pilypas, Harb, Buchler, Bachor, and Lam}}]{Hetet2007JoPBv40p221}
\bibinfo{author}{\bibfnamefont{G.}~\bibnamefont{Hétet}},
  \bibinfo{author}{\bibfnamefont{O.}~\bibnamefont{Glöckl}},
  \bibinfo{author}{\bibfnamefont{K.~A.} \bibnamefont{Pilypas}},
  \bibinfo{author}{\bibfnamefont{C.~C.} \bibnamefont{Harb}},
  \bibinfo{author}{\bibfnamefont{B.~C.} \bibnamefont{Buchler}},
  \bibinfo{author}{\bibfnamefont{H.-A.} \bibnamefont{Bachor}},
  \bibnamefont{and} \bibinfo{author}{\bibfnamefont{P.~K.} \bibnamefont{Lam}},
  \bibinfo{journal}{J. Phys. B} \textbf{\bibinfo{volume}{40}},
  \bibinfo{pages}{221} (\bibinfo{year}{2007}).

\bibitem[{\citenamefont{Predojevic et~al.}(2008)\citenamefont{Predojevic, Zhai,
  Caballero, and Mitchell}}]{Predojevic2008PRAv78p63820}
\bibinfo{author}{\bibfnamefont{A.}~\bibnamefont{Predojevic}},
  \bibinfo{author}{\bibfnamefont{Z.}~\bibnamefont{Zhai}},
  \bibinfo{author}{\bibfnamefont{J.~M.} \bibnamefont{Caballero}},
  \bibnamefont{and} \bibinfo{author}{\bibfnamefont{M.~W.}
  \bibnamefont{Mitchell}}, \bibinfo{journal}{Phys. Rev. A}
  \textbf{\bibinfo{volume}{78}}, \bibinfo{eid}{063820}
  (pages~\bibinfo{numpages}{6}) (\bibinfo{year}{2008}).

\bibitem[{\citenamefont{Appel et~al.}(2009)\citenamefont{Appel, Windpassinger,
  Oblak, Hoff, Kj{\ae}rgaard, and Polzik}}]{Appel2009PNASv106p10960}
\bibinfo{author}{\bibfnamefont{J.}~\bibnamefont{Appel}},
  \bibinfo{author}{\bibfnamefont{P.~J.} \bibnamefont{Windpassinger}},
  \bibinfo{author}{\bibfnamefont{D.}~\bibnamefont{Oblak}},
  \bibinfo{author}{\bibfnamefont{U.~B.} \bibnamefont{Hoff}},
  \bibinfo{author}{\bibfnamefont{N.}~\bibnamefont{Kj{\ae}rgaard}},
  \bibnamefont{and} \bibinfo{author}{\bibfnamefont{E.~S.}
  \bibnamefont{Polzik}}, \bibinfo{journal}{Proc. Nat. Ac. Science}
  \textbf{\bibinfo{volume}{106}}, \bibinfo{pages}{10960}
  (\bibinfo{year}{2009}).

\bibitem[{\citenamefont{Schleier-Smith
  et~al.}(2009)\citenamefont{Schleier-Smith, Leroux, and
  Vuletic}}]{Schleier-Smith2009Avp}
\bibinfo{author}{\bibfnamefont{M.~H.} \bibnamefont{Schleier-Smith}},
  \bibinfo{author}{\bibfnamefont{I.~D.} \bibnamefont{Leroux}},
  \bibnamefont{and} \bibinfo{author}{\bibfnamefont{V.}~\bibnamefont{Vuletic}},
  \bibinfo{journal}{ArXiv}  (\bibinfo{year}{2009}).

\bibitem[{\citenamefont{Geremia et~al.}(2004)\citenamefont{Geremia, Stockton,
  and Mabuchi}}]{Geremia2004Sv304p270}
\bibinfo{author}{\bibfnamefont{J.}~\bibnamefont{Geremia}},
  \bibinfo{author}{\bibfnamefont{J.}~\bibnamefont{Stockton}}, \bibnamefont{and}
  \bibinfo{author}{\bibfnamefont{H.}~\bibnamefont{Mabuchi}},
  \bibinfo{journal}{Science} \textbf{\bibinfo{volume}{304}},
  \bibinfo{pages}{270} (\bibinfo{year}{2004}).

\bibitem[{\citenamefont{Geremia et~al.}(2005)\citenamefont{Geremia, Stockton,
  and Mabuchi}}]{Geremia2005PRLv94p203002}
\bibinfo{author}{\bibfnamefont{J.}~\bibnamefont{Geremia}},
  \bibinfo{author}{\bibfnamefont{J.}~\bibnamefont{Stockton}}, \bibnamefont{and}
  \bibinfo{author}{\bibfnamefont{H.}~\bibnamefont{Mabuchi}},
  \bibinfo{journal}{Phys. Rev. Lett.} \textbf{\bibinfo{volume}{94}},
  \bibinfo{pages}{203002} (\bibinfo{year}{2005}).

\bibitem[{\citenamefont{Geremia et~al.}(2008)\citenamefont{Geremia, Stockton,
  and Mabuchi}}]{Geremia2008PRLvp1}
\bibinfo{author}{\bibfnamefont{J.}~\bibnamefont{Geremia}},
  \bibinfo{author}{\bibfnamefont{J.}~\bibnamefont{Stockton}}, \bibnamefont{and}
  \bibinfo{author}{\bibfnamefont{H.}~\bibnamefont{Mabuchi}},
  \bibinfo{journal}{Phys. Rev. Lett.} \textbf{\bibinfo{volume}{101}},
  \bibinfo{eid}{039902} (pages~\bibinfo{numpages}{1}) (\bibinfo{year}{2008}),
  \bibinfo{note}{\emph{erratum concerning the two preceding publications}}.

\bibitem[{\citenamefont{Kubasik et~al.}(2009)\citenamefont{Kubasik,
  Koschorreck, Napolitano, de~Echaniz, Crepaz, Eschner, Polzik, and
  Mitchell}}]{Kubasik2009PRAv79p43815}
\bibinfo{author}{\bibfnamefont{M.}~\bibnamefont{Kubasik}},
  \bibinfo{author}{\bibfnamefont{M.}~\bibnamefont{Koschorreck}},
  \bibinfo{author}{\bibfnamefont{M.}~\bibnamefont{Napolitano}},
  \bibinfo{author}{\bibfnamefont{S.~R.} \bibnamefont{de~Echaniz}},
  \bibinfo{author}{\bibfnamefont{H.}~\bibnamefont{Crepaz}},
  \bibinfo{author}{\bibfnamefont{J.}~\bibnamefont{Eschner}},
  \bibinfo{author}{\bibfnamefont{E.~S.} \bibnamefont{Polzik}},
  \bibnamefont{and} \bibinfo{author}{\bibfnamefont{M.~W.}
  \bibnamefont{Mitchell}}, \bibinfo{journal}{Phys. Rev. A}
  \textbf{\bibinfo{volume}{79}}, \bibinfo{pages}{043815}
  (\bibinfo{year}{2009}).

\bibitem[{\citenamefont{Madsen and M{\o}lmer}(2004)}]{Madsen2004PRAv70p52324}
\bibinfo{author}{\bibfnamefont{L.~B.} \bibnamefont{Madsen}} \bibnamefont{and}
  \bibinfo{author}{\bibfnamefont{K.}~\bibnamefont{M{\o}lmer}},
  \bibinfo{journal}{Phys. Rev. A} \textbf{\bibinfo{volume}{70}},
  \bibinfo{pages}{052324} (\bibinfo{year}{2004}).

\bibitem[{\citenamefont{Jauch and Rohrlich}(1976)}]{Jauch1976}
\bibinfo{author}{\bibfnamefont{J.}~\bibnamefont{Jauch}} \bibnamefont{and}
  \bibinfo{author}{\bibfnamefont{F.}~\bibnamefont{Rohrlich}},
  \emph{\bibinfo{title}{The Theory of Photons and Electrons}}
  (\bibinfo{publisher}{Springer, Berlin}, \bibinfo{year}{1976}).

\bibitem[{\citenamefont{Geremia et~al.}(2006)\citenamefont{Geremia, Stockton,
  and Mabuchi}}]{Geremia2006PRAv73p42112}
\bibinfo{author}{\bibfnamefont{J.}~\bibnamefont{Geremia}},
  \bibinfo{author}{\bibfnamefont{J.}~\bibnamefont{Stockton}}, \bibnamefont{and}
  \bibinfo{author}{\bibfnamefont{H.}~\bibnamefont{Mabuchi}},
  \bibinfo{journal}{Phys. Rev. A} \textbf{\bibinfo{volume}{73}},
  \bibinfo{pages}{042112} (\bibinfo{year}{2006}).

\bibitem[{\citenamefont{Lewandowski et~al.}(2003)\citenamefont{Lewandowski,
  Harber, Whitaker, and Cornell}}]{Lewandowski2003JoLTPv132p309}
\bibinfo{author}{\bibfnamefont{H.}~\bibnamefont{Lewandowski}},
  \bibinfo{author}{\bibfnamefont{D.}~\bibnamefont{Harber}},
  \bibinfo{author}{\bibfnamefont{D.}~\bibnamefont{Whitaker}}, \bibnamefont{and}
  \bibinfo{author}{\bibfnamefont{E.}~\bibnamefont{Cornell}},
  \bibinfo{journal}{J. Low Temp. Phys.} \textbf{\bibinfo{volume}{132}},
  \bibinfo{pages}{309} (\bibinfo{year}{2003}).

\bibitem[{\citenamefont{Ketterle et~al.}(1999)\citenamefont{Ketterle, Durfee,
  and Stamper-Kurn}}]{Ketterle1999PISoPFvp67}
\bibinfo{author}{\bibfnamefont{W.}~\bibnamefont{Ketterle}},
  \bibinfo{author}{\bibfnamefont{D.}~\bibnamefont{Durfee}}, \bibnamefont{and}
  \bibinfo{author}{\bibfnamefont{D.}~\bibnamefont{Stamper-Kurn}},
  \bibinfo{journal}{Proc. Int. School of Physics-Enrico Fermi}
  p.~\bibinfo{pages}{67} (\bibinfo{year}{1999}).

\bibitem[{\citenamefont{Windpassinger et~al.}(2009)\citenamefont{Windpassinger,
  Kubasik, Koschorreck, Boisen, Kjaergaard, Polzik, and
  Müller}}]{Windpassinger2009MSTv20p55301}
\bibinfo{author}{\bibfnamefont{P.~J.} \bibnamefont{Windpassinger}},
  \bibinfo{author}{\bibfnamefont{M.}~\bibnamefont{Kubasik}},
  \bibinfo{author}{\bibfnamefont{M.}~\bibnamefont{Koschorreck}},
  \bibinfo{author}{\bibfnamefont{A.}~\bibnamefont{Boisen}},
  \bibinfo{author}{\bibfnamefont{N.}~\bibnamefont{Kjaergaard}},
  \bibinfo{author}{\bibfnamefont{E.~S.} \bibnamefont{Polzik}},
  \bibnamefont{and} \bibinfo{author}{\bibfnamefont{J.~H.}
  \bibnamefont{Müller}}, \bibinfo{journal}{Meas. Sci. Technol.}
  \textbf{\bibinfo{volume}{20}}, \bibinfo{pages}{055301}
  (\bibinfo{year}{2009}).

\bibitem[{\citenamefont{Aigner et~al.}(2008)\citenamefont{Aigner, Della~Pietra,
  Japha, Entin-Wohlman, David, Salem, Folman, and
  Schmiedmayer}}]{Aigner2008Sv319p1226}
\bibinfo{author}{\bibfnamefont{S.}~\bibnamefont{Aigner}},
  \bibinfo{author}{\bibfnamefont{L.}~\bibnamefont{Della~Pietra}},
  \bibinfo{author}{\bibfnamefont{Y.}~\bibnamefont{Japha}},
  \bibinfo{author}{\bibfnamefont{O.}~\bibnamefont{Entin-Wohlman}},
  \bibinfo{author}{\bibfnamefont{T.}~\bibnamefont{David}},
  \bibinfo{author}{\bibfnamefont{R.}~\bibnamefont{Salem}},
  \bibinfo{author}{\bibfnamefont{R.}~\bibnamefont{Folman}}, \bibnamefont{and}
  \bibinfo{author}{\bibfnamefont{J.}~\bibnamefont{Schmiedmayer}},
  \bibinfo{journal}{Science} \textbf{\bibinfo{volume}{319}},
  \bibinfo{pages}{1226} (\bibinfo{year}{2008}).

\bibitem[{\citenamefont{de~Echaniz et~al.}(2008)\citenamefont{de~Echaniz,
  Koschorreck, Napolitano, Kubasik, and Mitchell}}]{Echaniz2008PRAv77p32316}
\bibinfo{author}{\bibfnamefont{S.~R.} \bibnamefont{de~Echaniz}},
  \bibinfo{author}{\bibfnamefont{M.}~\bibnamefont{Koschorreck}},
  \bibinfo{author}{\bibfnamefont{M.}~\bibnamefont{Napolitano}},
  \bibinfo{author}{\bibfnamefont{M.}~\bibnamefont{Kubasik}}, \bibnamefont{and}
  \bibinfo{author}{\bibfnamefont{M.~W.} \bibnamefont{Mitchell}},
  \bibinfo{journal}{Phys. Rev. A} \textbf{\bibinfo{volume}{77}},
  \bibinfo{pages}{032316} (\bibinfo{year}{2008}).

\end{thebibliography}

\end{document}